\documentclass[aps,prb,twocolumn,superscriptaddress,amsmath,amssymb]{revtex4}

\DeclareMathOperator{\Tr}{Tr}
\newcommand{\bra}[1]{\langle #1 \rvert}
\newcommand{\ket}[1]{\lvert #1 \rangle}

\begin{document}

\title{Comment on ``Quantum Monte Carlo scheme for frustrated Heisenberg antiferromagnets''}

\author{K.\ S.\ D.\ Beach}
\affiliation{Institut\,\,f\"ur\,\,theoretische\,\,Physik\,\,und\,\,Astrophysik,\,\,Universit\"at\,\,W\"urzburg,\,\,Am Hubland,\,\,D-97074\,\,W\"urzburg,\,\,Germany}
\affiliation{Department\,\,of\,\,Physics,\,\,University\,\,of\,\,Alberta,\,\,Edmonton,\,\,Alberta\,\,T6G 2G7,\,\,Canada}

\author{Matthieu Mambrini}
\affiliation{Laboratoire\,\,de\,\,Physique\,\,Th\'eorique,\,\,IRSAMC,\,\,CNRS,\,\,Universit\'e\,\,Paul\,\,Sabatier,\,\,31062\,\,Toulouse,\,\,France}

\author{Fabien Alet}
\affiliation{Laboratoire\,\,de\,\,Physique\,\,Th\'eorique,\,\,IRSAMC,\,\,CNRS,\,\,Universit\'e\,\,Paul\,\,Sabatier,\,\,31062\,\,Toulouse,\,\,France}

\begin{abstract}
Quantum Monte Carlo methods are sophisticated numerical techniques for
simulating interacting quantum systems. In some cases, however, they suffer from the notorious ``sign
problem'' and become too inefficient to be useful. A recent publication
[J. Wojtkiewicz, Phys. Rev. B {\bf 75}, 174421 (2007)] claims to have solved the
sign problem for a certain class of frustrated quantum spin systems
through the use of a bipartite valence bond basis. We show
in this Comment that the apparent positivity of the path integral
is due to a misconception about the resolution of the identity operator
in this basis, and that consequently the sign problem remains a severe obstacle for the simulation 
of frustrated quantum magnets.
\end{abstract}

\maketitle

Frustrated quantum antiferromagnets are of great interest, from both
experimental and theoretical points of view. A theoretical analytical
framework for understanding frustrated quantum magnets is still lacking, and
numerical simulations also turn out to be very difficult. In particular,
Quantum Monte Carlo (QMC) methods suffer from the ``sign problem,''
rendering QMC simulations of frustrated magnets completely inefficient.

In a recent publication,~\cite{Wojtkiewicz07} J.~Wojtkiewicz claims to have solved the
sign problem for a certain class of frustrated antiferromagnets, including
the important cases of the $J_1$--$J_2$ model and the pyrochlore lattice.
He proposes using a valence bond (VB) basis, rather than the
conventional $S_z$ basis, to formulate the path integral used in the QMC
framework. In this basis, the matrix elements involved are strictly positive,
and the sign problem seems to disappear. In this Comment, we point out that
the conventional route to constructing the path integral
(via the Trotter-Suzuki formula) is not suited to the bipartite VB basis used by Wojtkiewicz,
and that consequently the method proposed in Ref.~\onlinecite{Wojtkiewicz07} cannot work.

Let us first summarize the approach taken in Ref.~\onlinecite{Wojtkiewicz07}. The starting
point is to group terms in the Hamiltonian $H = \sum_P H_P$ according to the plaquette $P$
in which they act. The Trotter-Suzuki formula
[see Eq.~(1) in Ref.~\onlinecite{Wojtkiewicz07}], is used to decouple the potentially 
noncommuting terms appearing in the partition
function:
\begin{equation}
Z = \Tr e^{-\beta H} = \lim_{n\to\infty} \Tr \left[\prod_P e^{-\epsilon H_P}\right]^n.
\end{equation}
Here, $\epsilon=\beta/n$ is an infinitesimal imaginary-time step.
In the usual way, a representation of unity is introduced between each exponential factor,
leading to
\begin{equation} \label{EQ:Z}
Z = \sum_{\{\alpha\}} \prod_{i=1}^n \prod_P \bra{\alpha_{P,i}}e^{-\epsilon H_P}\ket{\alpha_{P,i+1}},
\end{equation}
where the notation implies $\ket{\alpha_{P,n+1}} = \ket{\alpha_{P,1}}$. This amounts to introducing 
a new set of states at each plaquette and at each time slice.

Implicit in Eq.~\eqref{EQ:Z} is the existence of 
a simple resolution of the identity operator,
\begin{equation}
\label{Id}
\hat{1} = \sum_{\alpha} \ket{\alpha} \bra{\alpha},
\end{equation}
in the basis $\{ \alpha \}$ used to perform the simulations. Taking this
for granted, Ref.~\onlinecite{Wojtkiewicz07} calculates all possible matrix
elements $\bra{\alpha} e^{-\epsilon H_P} \ket{\alpha'}$
(with $\ket{\alpha}$ and $\ket{\alpha'}$ in the \emph{bipartite} VB basis) and
shows that they are all positive. Wojtkiewicz then concludes that a QMC process 
without sign problem can be devised with this decomposition.

The essence of this Comment is simple: we point out that Eq.~\eqref{Id}
does {\it not} hold in the bipartite VB basis considered for the
calculations. The correct resolution of identity reintroduces the sign
problem in the Trotter-Suzuki decomposition of the partition function and
renders the scheme proposed in Ref.~\onlinecite{Wojtkiewicz07} useless.

In the bipartite valence bond (BVB) basis, the lattice is 
decomposed into two sets (A and B) of sites, and
the two spins forming a singlet necessarily belong to different sets:
$[a,b]=(\ket{\uparrow_a   \downarrow_b}-\ket{\downarrow_a
  \uparrow_b})/\sqrt{2}$. The bipartite VB basis is overcomplete, albeit
smaller than the full valence bond (FVB) basis, where there are no restrictions on the
constituent spins of the singlet. Since VB bases are nonorthogonal, the
most general form of the identity operator is
\begin{equation}
\label{Id2}
\hat{1} = \sum_{\alpha,\alpha'} \ket{\alpha}1_{\alpha,\alpha'}\bra{\alpha'}.
\end{equation}
This means
that one should also consider the matrix elements of $1_{\alpha,\alpha'}$ at
the QMC sampling level, and in particular ensure that they are all positive in
order to avoid the sign problem.

The massive overcompleteness of the basis offers multiple different solutions for
$1_{\alpha,\alpha'}$. One of the possible choices in the FVB basis is particularly
simple. Indeed, for a system of $2N$ spins, $1_{\alpha,\alpha'}$ 
can take the diagonal form $\frac{2^N}{(N+1)!}\delta_{\alpha,\alpha'}$ [see Eq.~(27) in
Ref.~\onlinecite{Beach06}]. But, such a simple resolution is not possible if the
basis consists of AB bonds only~\cite{note-minus} (BVB basis).

To see this explicitly, consider the four-site ($N=2$) system, which admits
three VB states---two bipartite, $[1,2][3,4]$ and $[1,4][3,2]$,
and one nonbipartite, $[1,3][2,4]$.
The representation of the identity operator has only positive, diagonal elements in the full VB basis
but negative, offdiagonal ones in the bipartite basis:
\begin{align}
 \bigl( 1_{\alpha,\alpha'}\bigr)_\mathrm{FVB} &=
		\frac{2}{3}
                     \begin{pmatrix}
                      1 & 0 & 0 \\
                      0 & 1 & 0 \\
                      0 & 0 & 1
                     \end{pmatrix} \\
\bigl( 1_{\alpha,\alpha'}\bigr)_\mathrm{BVB} &=
		\frac{2}{3}
                      \begin{pmatrix}
                       2 & -1  \\
                       -1 & 2  
                      \end{pmatrix}
\end{align}

Let us comment on the general $N$ case: the matrix $1_{\alpha, \alpha'}$ 
is the pseudo-inverse of the positive semidefinite overlap matrix 
$O_{\alpha,\alpha'} = \langle \alpha | \alpha'\rangle$ 
(when only AB bonds are allowed). Since $O_{\alpha,\alpha'}$ has 
only positive elements, the positivity of all off-diagonal elements of 
$1_{\alpha,\alpha'}$ would require that all diagonal elements
be strictly negative. This is in condradiction with the positivity of the 
traces of both $O_{\alpha,\alpha'}$ and $1_{\alpha, \alpha'}$. 
Therefore, in the basis of bipartite bonds, $1_{\alpha, \alpha'}$ must 
contain at least one strictly negative, off-diagonal element. Indeed, 
one can establish that $1_{\alpha, \alpha'}$ is dense in such elements.

Alternatively, one could imagine formulating the QMC using the full VB basis, which has a simpler
resolution of identity. Unfortunately, the sign problem then appears in the
Hamiltonian matrix elements. In short, we are left with two ways to organize the path integral, one involving a sum over a VB basis restricted to AB bonds only,
\begin{equation}
Z = \sum_{\{\alpha\}}^{\text{BVB}} \cdots\underbrace{\bra{\alpha}e^{-\epsilon H_P}\ket{\alpha'}}_{\text{positive}}\underbrace{1_{\alpha',\alpha''}}_{\text{signed}}
\underbrace{\bra{\alpha''}e^{-\epsilon H_P}\ket{\alpha'''}}_{\text{positive}}\cdots,
\end{equation}
and one involving an unrestricted sum,
\begin{equation}
Z = \sum_{\{\alpha\}}^{\text{FVB}}\cdots\underbrace{\bra{\alpha}e^{-\epsilon H_P}\ket{\alpha'}}_{\text{signed}}
\underbrace{\bra{\alpha'}e^{-\epsilon H_P}\ket{\alpha''}}_{\text{signed}}\cdots
\end{equation}
In neither case is the sign eliminated.
In the bipartite VB basis, the matrix elements $\bra{\alpha}e^{-\epsilon H_P}\ket{\alpha'}$ are strictly
positive, but the identity matrix elements $1_{\alpha,\alpha'}$ are not. In the full VB basis, the situation
is reversed: $1_{\alpha,\alpha'} \sim \delta_{\alpha,\alpha'} > 0$, but $\bra{\alpha}e^{-\epsilon H_P}\ket{\alpha'}$
is no longer strictly positive. Thus, the choice of basis amounts to nothing more than
shuffling the sign problem from one location to another.
With these considerations in mind, it is clear that the Trotter-Suzuki 
decomposition cannot be performed as envisioned in Ref.~\onlinecite{Wojtkiewicz07},
and the QMC scheme proposed therein breaks down.

In conclusion, the VB basis can be very useful in QMC simulations, but a workable algorithm cannot be
organized around inserting an overcomplete set of states between time slices in a 
Trotter-Suzuki scheme. A better approach is to project down from a trial state
using the power method, relying on the property that Heisenberg spin interactions
induce simple rearrangements of valence bond configurations.~\cite{Sandvik05}
In this way, spin models with nonfrustrating interactions of arbitrary
order can be treated. Grouping of terms (along the lines of 
the plaquette decomposition of Ref.~\onlinecite{Wojtkiewicz07}) combined with a fixed-node
approximation~\cite{Bemmel94} might partially alleviate the sign problem 
in models with frustrating interactions
and lead to more efficient QMC sampling,~\cite{Beachxx} 
but we do not expect a simple full solution to the sign problem (such as the one
claimed in Ref.~\onlinecite{Wojtkiewicz07}) to emerge from this approach.

Finally, it is worth noting that the \emph{overlap calculations} reported in 
Ref.~\onlinecite{Wojtkiewicz07} are certainly correct and may yet prove useful in the
evaluation of variational valence bond wavefunctions, 
following the method of Liang, Dou\c{c}ot, and Anderson\cite{Liang88} and its refinements.~\cite{Lou06,Beach07}

We thank Sylvain Capponi and Anders Sandvik for fruitful discussions and
J.\ Wojtkiewicz for helpful correspondence. This research is supported by the
French ANR program (MM and FA), the Alexander von Humboldt foundation (KB), and the
French-German Procope exchange program.\\


\begin{thebibliography}{99}
\bibitem{note-minus}
This can be traced back to the minus sign existing in the
overcompleteness relation $[a_1,a_2][b_1,b_2] = [a_1,b_1][a_2,b_2] -
[a_1,b_2][a_2,b_1]$ that allows one to systematically eliminate all AA/BB bonds
when going from the FVB basis to the BVB one.
\end{thebibliography}
\end{document}